\documentclass[
    final            % use final for the camera ready runs
  ]
  {aipproc}

\layoutstyle{8x11single}

\begin{document}

\title{Low background techniques for the Borexino nylon vessels}


\author{Andrea Pocar}{
  address={Physics Department, Stanford University, Stanford, CA  94305  USA}
}

%%\author{<author2>}{
%%  address={<common address for author2 and author3>}
%%}

%%\author{<author3>}{
%%  address={<common address for author2 and author3>}
%% ,altaddress={<author1 address>} % additional visiting address
%%}

\begin{abstract}
Borexino is an organic liquid scintillator underground detector for low energy solar neutrinos. The experiment has to satisfy extremely stringent low background requirements. The thin nylon spherical scintillator containment vessel has to meet cleanliness and low radioactivity levels second only, within the detector, to the scintillator itself. Overall, the background from the vessel in the fiducial volume of the detector must be kept at the level of one event per day or better. The requirements, design choices, results from laboratory tests, and fabrication techniques that have been adopted to meet this goal are presented. Details of the precautions taken during the installation of the vessels inside the Borexino detector are also discussed.
\end{abstract}

\maketitle

%%%%%%%%%%%%%%%%%%%%%%%%%%%%%%%%%%%%%%%%%%%%
%% MAINMATTER
%%%%%%%%%%%%%%%%%%%%%%%%%%%%%%%%%%%%%%%%%%%%

\section{The Borexino Detector}
Borexino is a solar neutrino experiment approaching completion at the underground Gran Sasso national laboratories in central Italy \cite{bx-scitech:2002} . It is designed to detect low energy solar neutrinos ($^7$Be, pep) as well as antinuetrinos emitted in beta decays in the earth's crust \cite{rothschild:1998} and neutrinos from possible supernov\ae~in our galaxy \cite{bx-sn:2002}.

Borexino will detect neutrinos that elastic scatter off the electrons of its 300 ton organic scintillator target. The inner 100 tons are designed to have minimal background in the enrgy observation window for solar neutrinos (between 250 keV and 1.5 MeV). The scintillator is contained inside an 8.5 meter diameter spherical balloon-like nylon vessel (Inner Vessel, IV). The scintillation light flash produced by recoiling electrons will be detected by more than 2200 photomultiplier tubes (PMTs) aimed at the scintillator volume and mounted on a stainless steel sphere (SSS), concentric with the IV and 13.7 meters in diameter. A second nylon vessel (Outer Vessel, OV), concentric and essentially identical to the IV but with a diameter of 11 meters is placed between the IV and the PMTs; its role is to prevent radon emanated by the PMTs and the supporting steel sphere from diffusing into the core of the detector. The volume inside the SSS in then divided into three regions: these are, from the inside out, the Inner Vessel, the Inner Buffer (IB), and the Outer Buffer (OB). Each vessel has two sets of low activity ropes that carry the buoyant loads due to density differences between the volumes and temperature gradients.

\begin{figure}
  \includegraphics[height=.35\textheight]{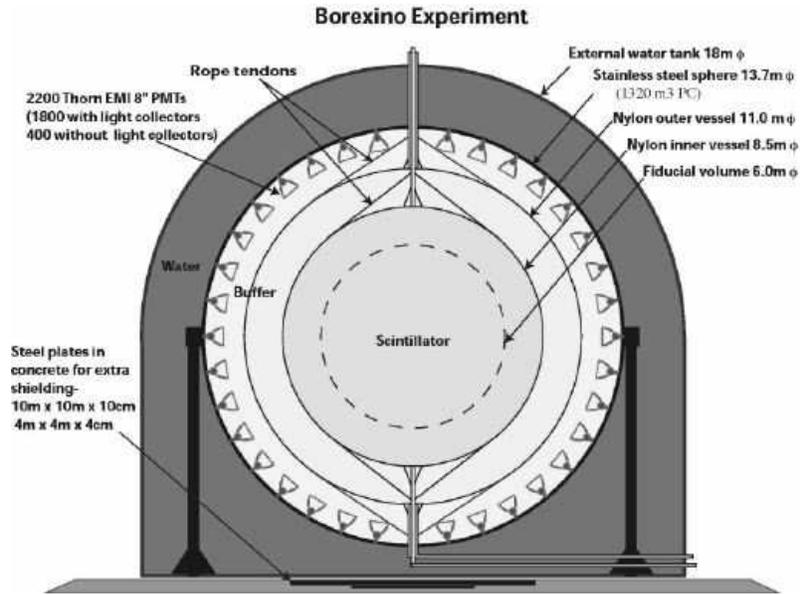}
  \caption{The Borexino detector. The active scintillator target is contained in a transparent, spherical nylon vessel, 125 $\mu$m thick. The low background requirements for the scintillator and the nylon vessel are extremely stringent, having to contribute on the order of one event per day in the energy window for neutrino detection below 1 MeV.}
  \label{f:bx}
\end{figure}

The scintillator is a solution of 1.5 g/l dyphenil oxazole (PPO) in 1,2,4-trimethylbenzene (pseudocumene). The volume between the SSS and the IV is filled with a solution of dimethylphthalate (DMP) in pseudocumene. This volume of fluid acts as a buffer for background radiation from the SSS and PMTs and the DMP quenches the light such radiation would generate at high rate in the detector. The choice of scintillator and buffer solutions with quasi zero buoyancy allowed the vessels to be extremely thin thus limiting their total mass. As explained in the following, this was one of many necessary steps in order to minimize the radioactive contamination close to the active region of the detector.
Ultra pure water contained in a tank around the SSS provides shielding for external neutrons from the experimental hall and is instrumented with PMTs to detect \v{C}erenkov light from muons that cross it. A detailed sketch of the Borexino detector is shown in fig.~\ref{f:bx}

\subsection{The nylon vessels}
The Borexino nylon vessels are two concentric spherical balloons, with diameters of 8.5 and 11 meters respectively. They are made out of a thin nylon membrane, just 125 $\mu$m thick. Nylon was chosen because it is transparent and optically clear, chemically inert in pseudocumene, radioactively very pure, and presents low permeation for gases (especially the radioactive noble gasses $^{39}$Ar, $^{85}$Kr, and $^{222}$Rn~\cite{cadonati-tesi:2001, wojcik:2000, wojcik:2004}). The choice of a thin film was driven by the effort of minimizing the amount of material close to the scintillator. An extensive treatment of the properties of the nylon membrane can be found in~\cite{cadonati-tesi:2001}.
At the poles, the vessels are connected to a set of pipes that carry the scintillator and buffer fluids into the IV, IB, and OB.

The following sections describe the low radioactivity requirements for the nylon vessels and the details of their design and fabrication explicitly meant to enhance radio-purity and cleanliness. Low radioactivity techniques employed for all the Borexino components are reported in \cite{bx-lowrad:2002}.

\section{Low radioactivity aspects of the Borexino nylon vessels}
Providing an ultra pure container for the 300 ton scintillator target is arguably the hardest challenge for the Borexino nylon vessel. The first natural step to take is to minimize their mass. As mentioned above, the quasi-zero buoyancy design between scintillator and buffer fluids made it possible to use 125 $\mu$m thick film, yielding a mass of the IV envelope of less than 50 kg. A strict material selection and screening was nonetheless necessary. Gamma-rays emitted at the IV by isotopes in the $^{238}$U and $^{232}$Th natural radioactive chains (especially $^{214}$Bi and $^{208}$Tl) and by $^{40}$K  (naturally present in potassium at the 0.01\% level) are a serious potential background (the attenuation length for 1 MeV $\gamma$ rays in pseudocumene is $\sim$25 cm). Last but not least, contamination from particulate and radon exposure during fabrication and installation inside the detector had to be kept at an absolute minimum. Indeed, whatever deposits on the inner surface of the IV can end up directly into the scintillator. In particular, $^{210}$Pb (32 year mean life) accumulation following radon exposure yields $^{210}$Bi (1.17 MeV $\beta$ decay) and $^{210}$Po (5.3 MeV $\alpha$ decay), both extremely dangerous backgrounds for Borexino. For particulate contamination, assuming its $^{238}$U and $^{232}$Th concentration to be 1 ppt, no more than 3 mg were allowed on the entire inner surface of the IV.

The thin film design made it possible to build the vessels as folded stacks on tables inside a clean room. Contamination from radon daughters was a main concern: a straightforward estimate made with somewhat pessimistic assumptions on the dynamics of radon daughters says that a normal fabrication procedure (in a clean room with a typical radon activity of 30 Bq/m$^3$) could lead to backgrounds three to four orders of magnitude too high inside the detector from this source \cite{pocar-tesi:2003, leung:2004}. Several aspects of the fabrication technique, illustrated below, were driven by the need to minimize the exposure of the vessel surfaces. Strict cleanliness standards and clean room policies were adopted. In addition, the clean room was specifically constructed to accommodate a radon filter for its supply air, which further reduced the radon contamination of the vessels. The filter and laboratory-scale studies of radon contamination of nylon surfaces are specifically discussed below.

\begin{figure}
  \includegraphics[height=.37\textheight]{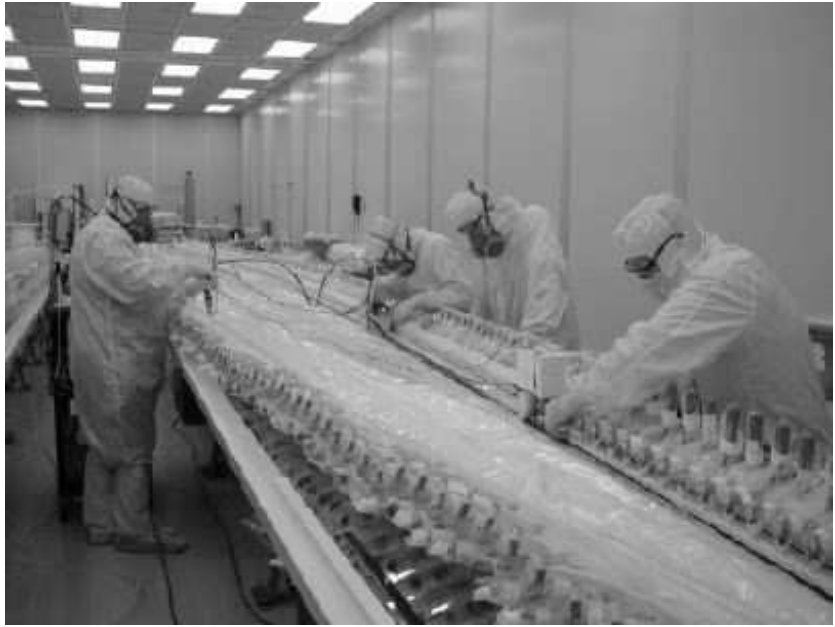}
  \caption{Gluing one of the joints of the IV. The large size of the clean room can be appreciated, as well as the effort made to cover all surfaces of the vessel not being bonded and to individually bag the clamps (used to apply pressure to the bonded panels during glue curing) to prevent particulate contamination of the vessel.}
  \label{f:joint}
\end{figure}

\paragraph{The clean room}
The Princeton clean room for nylon vessel fabrication is approximately $22\times6.5\times4.5$ meters in size (fig. \ref{f:joint}). It is certified to class 100, and was measured to be around class 10 when unperturbed (the class is the number of particles with diameter $\geq 5\,\mu$m per cubic foot of air).  Recirculation time of the air through HEPA filters is about 30 seconds. The water for humidification ($\sim$50\% RH) was aged for about 100 days for radon reduction.
The clean room was designed to be fed with low-radon air supplied by a radon filter.  This requires the clean room system to be leak-tight in order to prevent back-diffusion of radon and to reduce the amount of radon-free air needed to give the room the needed over-pressure.  The radon filter delivered $\sim$85~m$^3$/h of radon-scrubbed air during vessel construction and is described more in detail below. More details on the clean room are found in \cite{pocar-tesi:2003}, while the problem of radon contamination of surfaces is addressed in \cite{pocar-tesi:2003, leung:2004}.

\paragraph{Film selection and production}
Two raw nylons ("Capron" by Allied Signal/Honeywell and "Sniamid" by Nyltech) were selected among many due to their high purity in the pellet state. The measured concentrations of $^{238}$U and $^{232}$Th were $\sim$1 ppt for both, with Capron being slightly less contaminated and thus the first choice. For both, the concentration of $^{\rm nat}$K was < 10 ppb. For details on these measurements and for an explanation on the differences between the two nylons, refer to \cite{cadonati-tesi:2001} and references therein. 

Two extrusions campaigns were carried out: one at Honeywell with Capron and the other at MF-Folien with Sniamid. For both, great care was put in keeping the area around the extruding machines clean. After extrusion, the film was precision cleaned at CleanFilm with a non-contact technique where particles were first ultrasonically disrupted from the surface and then removed by negative pressure. Its surface was certified to level 25 Mil. Std. 1246C (less than one particle 25 $\mu$m in diameter per square foot \cite{milstd1246c}).

Only after the OV, made out of Capron, was completed, radon emanation results became available for both films. This was the ultimate test of the final extruded film after it underwent the cleaning process. It also measured the concentration of $^{226}$Ra rather than $^{238}$U higher in the decay chain, automatically including the possibility that the two isotopes are not in equilibrium. Capron turned out to be about an order of magnitude more contaminated than Sniamid, with the latter having a $^{226}$Ra activity of 16 $\mu$Bq/kg \cite{zuzel:2003}, the lowest ever measured in any material (this translates to 1.3 ppt $^{238}$U assuming secular equilibrium, or approximately 1000 $^{226}$Ra atoms per gram of nylon!). The technique adopted for the radon emanation measurement from the film was able to separate bulk and surface contaminations by using the humidity dependance of the diffusion coefficient of radon through nylon. A $^{226}$Ra surface contamination <0.8 $\mu$Bq/m$^2$ could be set, thus proving the effectiveness of the surface cleaning procedure \cite{zuzel:2003} .

\paragraph{Other components}
The material used to build the polar "end caps" and the pipes for the scintillator were also carefully screened. The section of the IV pipe closest ($\sim$1 m) to the vessel is made of nylon, while stainless steel was used for the sections at larger radii. Nylon is in fact less dense and much less radioactive. In addition, extruded nylon parts were found to be more radio-pure than cast ones. The mass of the end caps was kept to a minimum (about 15 kg), and the activity of the material presented $\sim$50 ppt concentrations of $^{238}$U and $^{232}$Th and $\sim$500 ppb of $^{\rm nat}$K. Nylon and stainless steel sections were precision cleaned at Astropak and their surfaces also certified to level 25 Mil. Std. 1246C \cite{milstd1246c}.
Most of the ropes considered for the Borexino vessels were found to have a high potassium content. Those selected for Borexino are made out of ultra high density polyethylene (UHDPE) and have a $^{\rm nat}$K contamination of approximately 1 ppm.

\begin{figure}
  \includegraphics[height=.275\textheight]{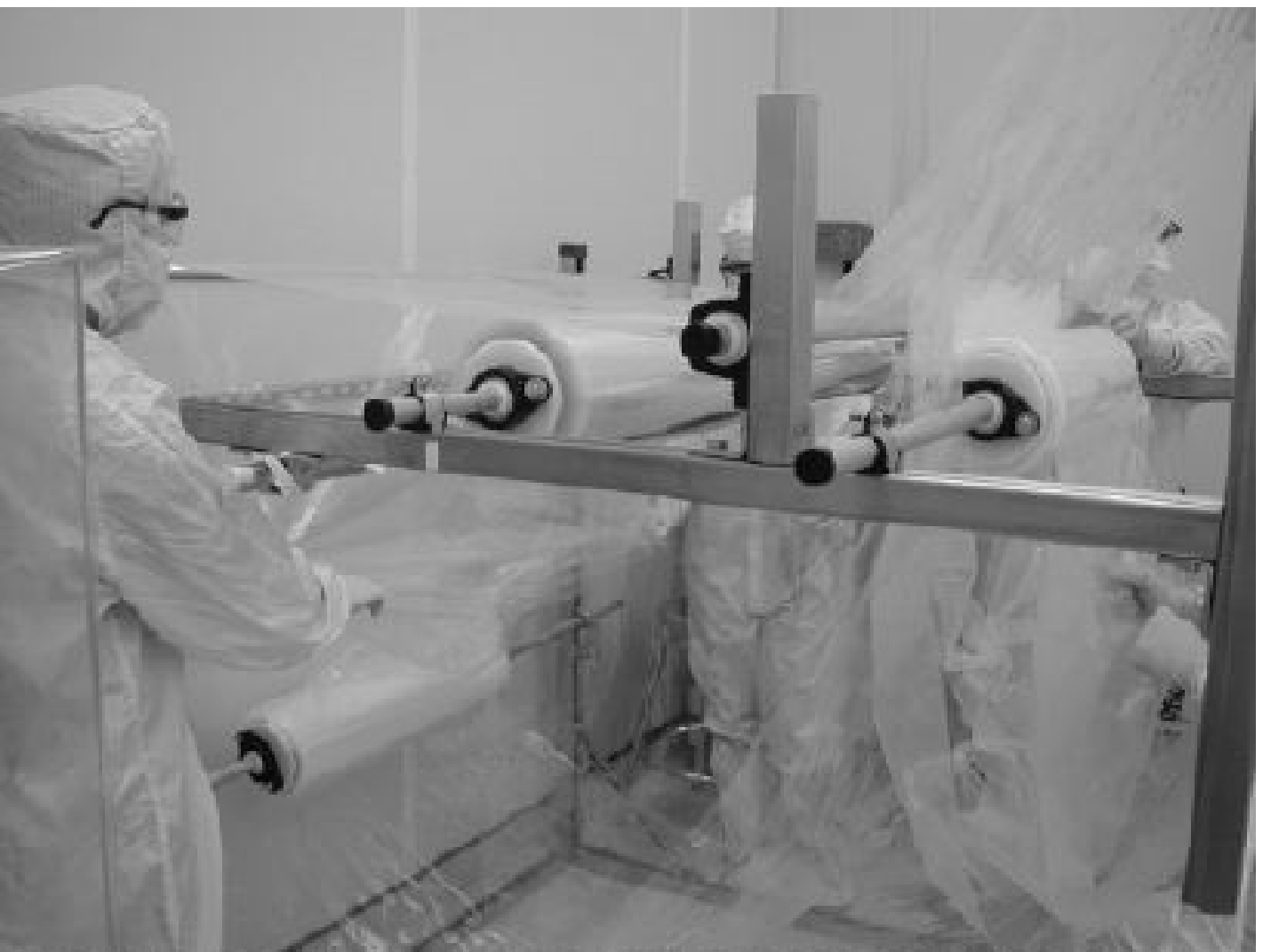}
  \includegraphics[height=.275\textheight]{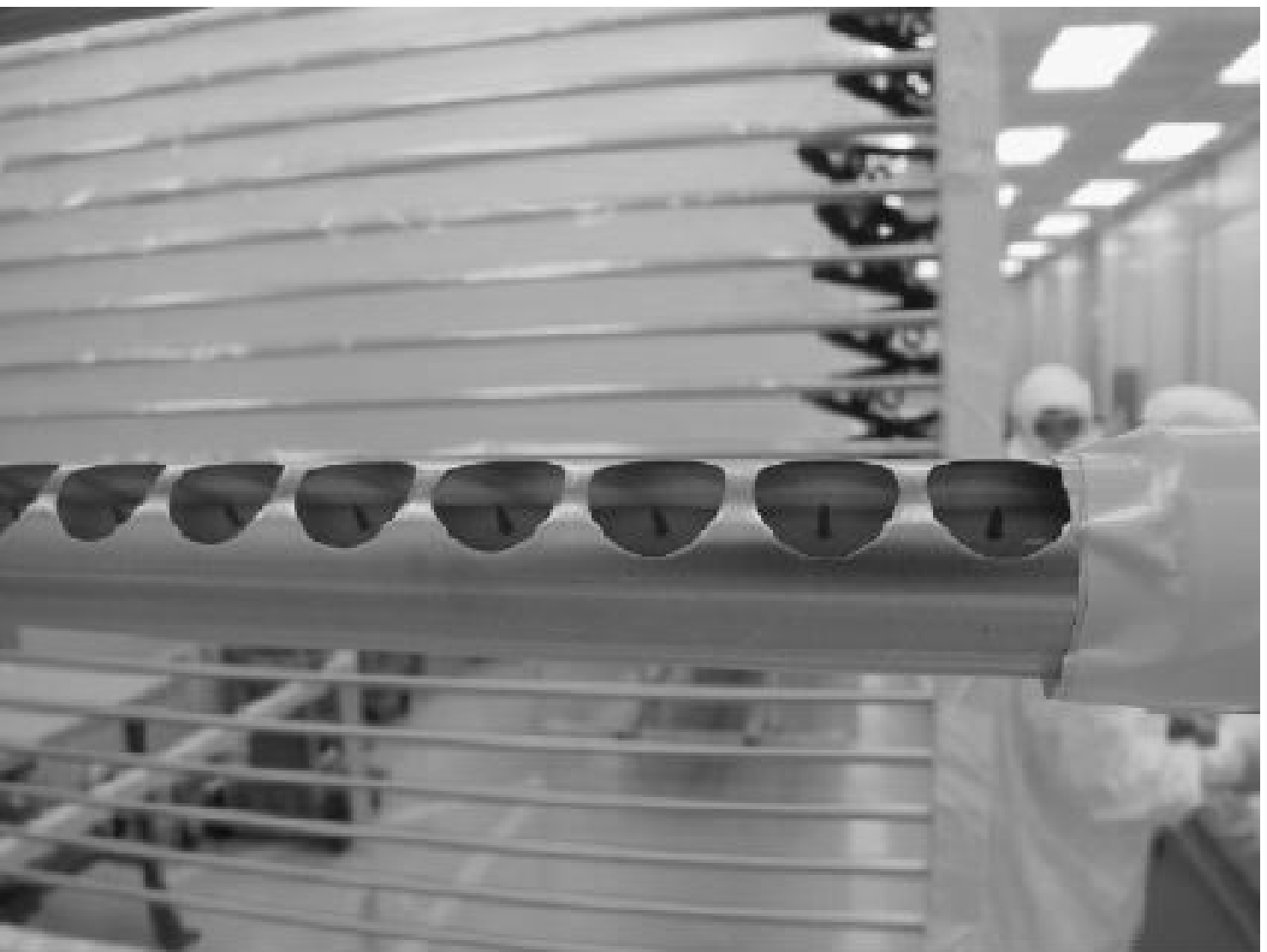} 
  \caption{The nylon film was immediately sandwiched between sheets of clean nylon film to protected it from radon daughter deposition and residual particulate (left). To minimize radon daughter contamination during this operation, the nylon membrane was discharged from the accumulated static charge due to friction by two high voltage de-ionizing bars (right).}
  \label{f:cover}
\end{figure}

\paragraph{Vessel fabrication}
Fabrication of the nylon vessels started in the summer of 2001 and lasted for a year, until August 2002. Film was first unrolled and let acclimatize to the humidity of the clean room to their equilibrium elongation (few days). Then almond-shaped panels were cut out (36 for the IV, 40 for the OV): glued together, these would define a sphere when inflated. During these steps, a big effort was put into keeping the nylon film clean by protecting it from residual particulate and deposition of radon daughters. The nylon coming off the spool was immediately sandwiched between two layers of extremely clean (level 25 Mil. Std. surface cleanliness \cite{milstd1246c}), 25 $\mu$m-thick nylon sheets (fig. \ref{f:cover}, left). In order to prevent deposition of radon daughter ions during this step where the dry film typically accumulates static charge on its surface, two high voltage de-ionizing bars were employed (fig. \ref{f:cover}, right).

\begin{figure}[b]
  \includegraphics[height=.1\textheight]{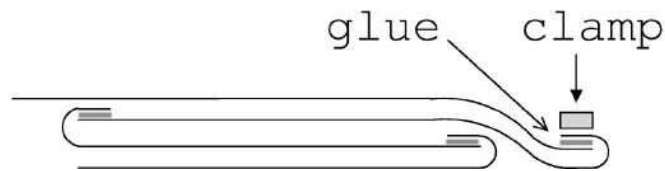}
  \caption{The nylon vessels were built in accordion-shaped stacks which have the advantage of being self covering for the panels below the top one. This geometry also suppresses radon diffusion by reducing the possible air gaps in contact with the nylon surface.}
  \label{f:gluejoint}
\end{figure}

The panels were glued together on tables in a flat accordion-shaped stack (fig. \ref{f:gluejoint}). Each panel is bonded to the one below along a folded edge. Glue is applied on both sides of the joint and pressure is applied for several hours to obtain strong adhesion. During this step, all panels are kept covered on both sides and only the area being glue is directly exposed to the clean room atmosphere for the time needed. More details on this procedure are found in \cite{pocar-tesi:2003}. The stack geometry has the advantage of being efficiently self covering, with the top panel protecting all the underlying ones. The air gaps between panels are extremely small so that radon diffusion into the stack is highly suppressed. The whole stack was also kept covered with additional layers of clean film and a sheet of thick, aluminized film during down time to reduce radon exposure.

\begin{figure}
  \includegraphics[height=.27\textheight]{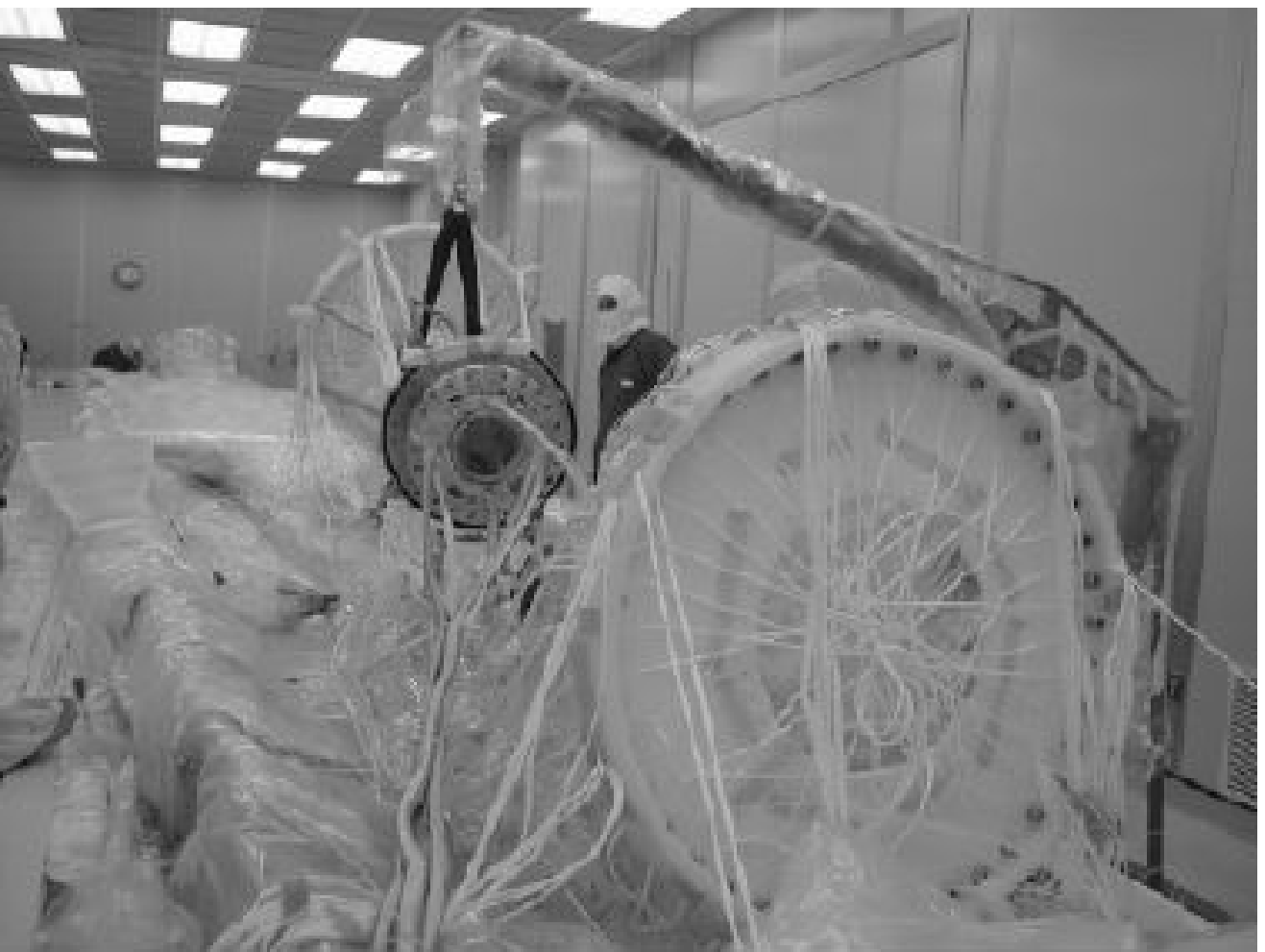}
  \includegraphics[height=.27\textheight]{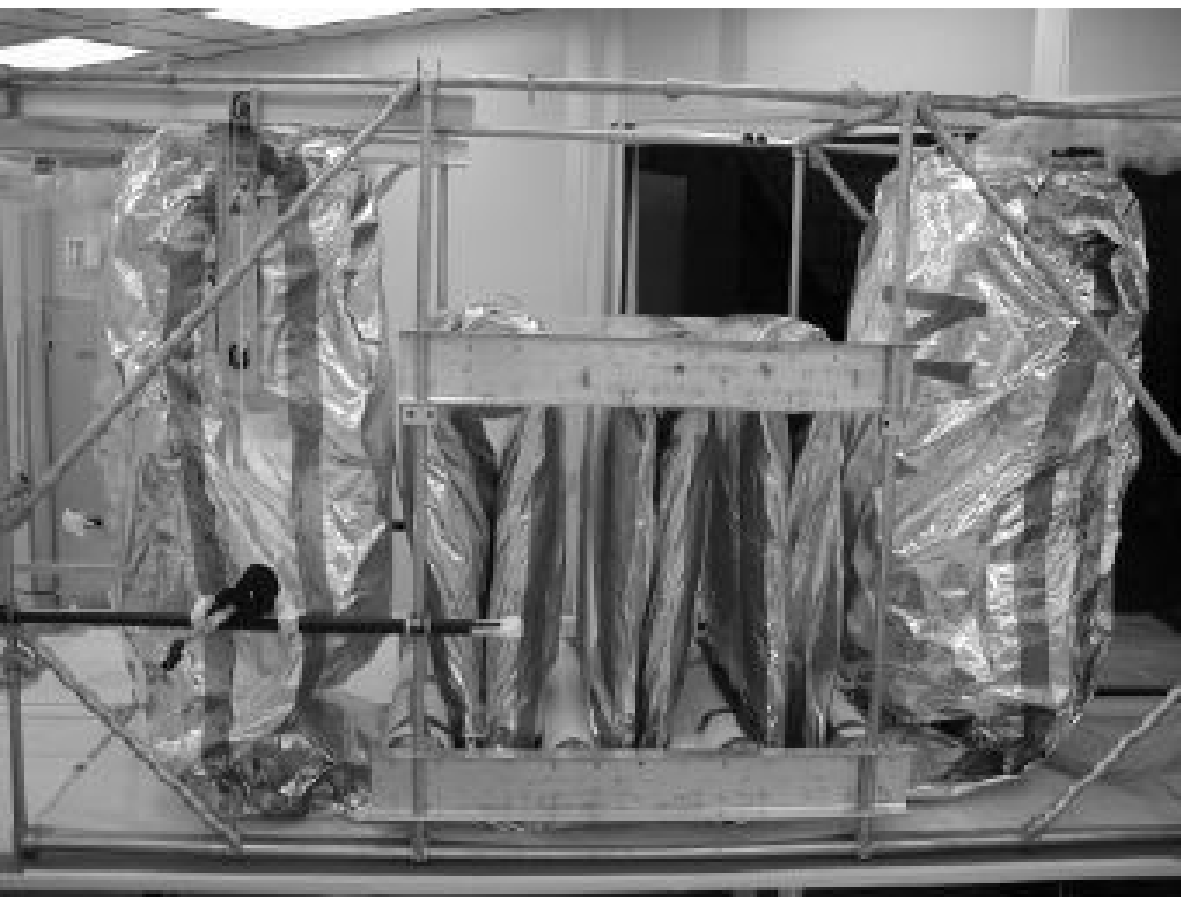} 
  \caption{Left: the IV during the nesting operation into the OV. The presence of a very leak-tight shroud in a single package assembled in a radon-suppressed clean room environment adds a layer of protection for maintaining the cleanliness of the IV surface. Right: the completed nylon vessel system (IV nested inside the OV) is wrapped with several layers of clean film of various thicknesses and sealed in a thick aluminized foil to minimize radon diffusion.}
  \label{f:nesting}
\end{figure}

The IV was then sealed with the polar end regions and leak checked (see \cite{pocar-tesi:2003} for details). The vessel assembly sequence foresaw nesting the IV into the OV into a single package. This operation took place in the clean room (fig. \ref{f:nesting}, left). The presence of a very leak tight envelope around the IV, assembled in a clean room environment added a layer of protection for keeping the surface of the IV clean. Once the vessels were nested into each other, the OV was sealed along the last open seam and the assembly was packaged and folded for shipping. Several layers of clean film of various thicknesses were used. The package was finally sealed in a thick layer of aluminized film against mechanical damage and radon diffusion (fig. \ref{f:nesting}, right).

\section{The radon filter}
\label{ss:scrubber}
\begin{figure}
  \includegraphics[height=.32\textheight]{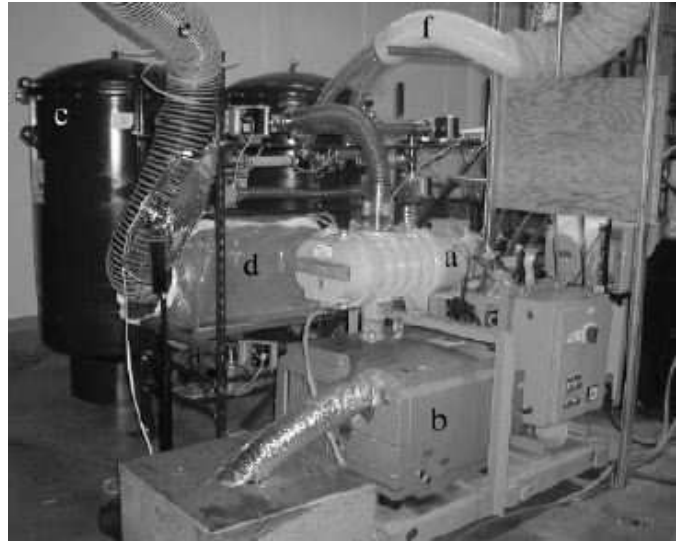}
  \caption{The radon filtered that supplied the entire make up air flow to the vessel assembly clean room. The device operated continuously for the whole year of vessel fabrication. The main components are labelled: a,b) vacuum pump (roots blower + backing pump, respectively), c) charcoal columns, d) hepa filter at the output, mounted in a frame together with the valve manifold, e) duct to the clean room, f) column repressurizarion line. }
  \label{f:scrubber}
\end{figure}

A radon filter of novel design was built to supply radon-suppressed air to the clean room for nylon vessel assembly. The filter is based on room temperature vacuum-swing adsorption on activated charcoal. It operated continuously during the whole year of vessel assembly, radon-scrubbing the entire flow of make up air required by the clean room. Such flow averaged around 85 m$^3$/h ($\sim$ 25 liters/minute). The device is shown in fig. \ref{f:scrubber} and is described in detail in \cite{pocar-tesi:2003}.  The main results from the radon filter and the prototypes on which it was founded are illustrated in the next few paragraphs.\\

\paragraph{Principle of operation}
Vacuum-swing adsorption (VSA) is conceptually identical to pressure-swing adsorption (PSA), a widely used process for gas separation \cite{ruthven-psa:1994, yang:1987}. Adsorption on a surface is a bond between an atom or molecule and molecular sites on the surface. It is, in its simplest form, a dipole-dipole (Van der Waals) electromagnetic interaction whose strength falls as the distance to the sixth power. It's strength depends on both the species to be adsorbed and the surface material used. In the case of activated charcoal, the adsorption binding energy is stronger for a radon atom than for the other components of air (oxygen, nitrogen, argon, ...). Thus, when traveling through a charcoal bed (column), radon is effectively slower than other species in the flow, and the air emerging from the output is radon-depleted. 

\begin{figure}[b]
  \includegraphics[height=.3\textheight]{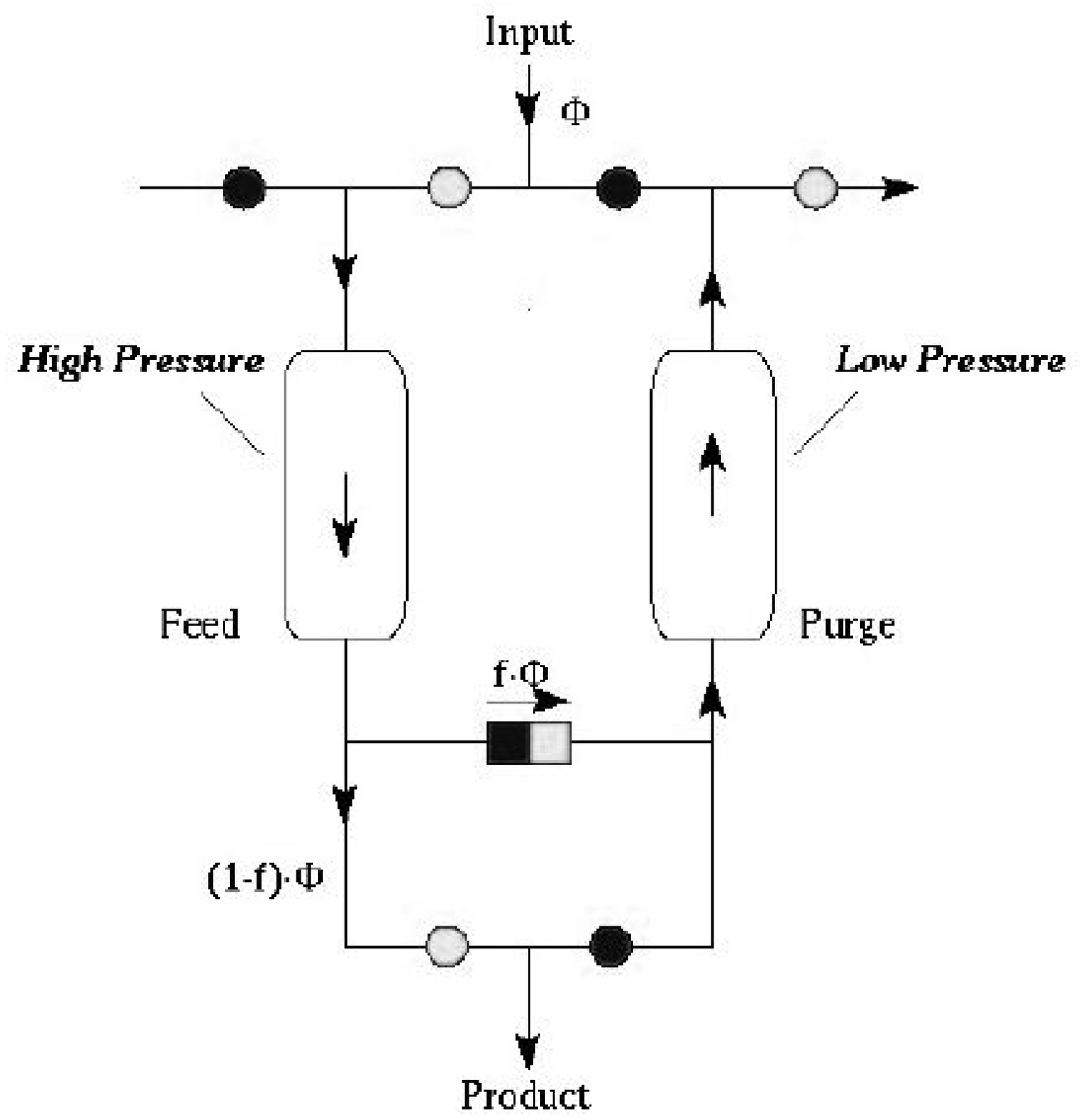}
  \includegraphics[height=.29\textheight]{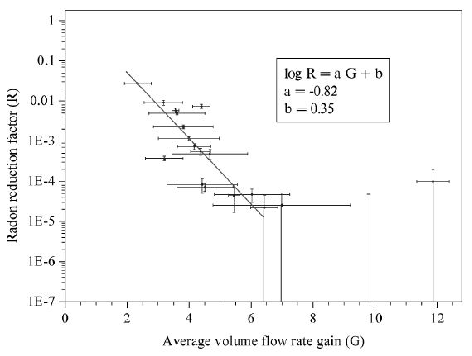}
  \caption{Schematic representation of a pressure-swing system. A fraction $f$ of the clean product $\Phi$ from the feed column is expanded and sent through the other column in counterflow. Light an dark dots represent open and closed valves. {\bf Right}: radon reductions ($R$) obtained with a small-scale VSA system on activated charcoal. Data are obtained contaminating the input air stream with a radon source. The line represents the best fit to the data (that show a non-negligible spread). The gain $G$ is expressed as an average to account for pressure drops that are more significant across the purged column.}
  \label{f:psa}
\end{figure}

A pressure (vacuum)-swing system is composed of at least two identical charcoal columns which undergo a cycle of feed and purge (fig. \ref{f:psa}, left). One of the columns filters the full air flow, while a small fraction of the clean output is expanded at lower pressure and fed in counter flow through the other column. This high velocity purge flow collects the impurities (radon in this case) accumulated during the previous cycle when said column was used as the filter, and discharges them. At the end of this stage, the purged column is clean and ready to work as a filter again, while the other is loaded with impurities and ready to undergo a purge cycle. The time after which the two columns switch roles depends on the application and degree of purity required. A vacuum swing system uses feed at atmospheric and purge at sub-atmospheric pressure. Swing systems thus provide continuos supply of filtered gas stream by using multiple columns and continuous inline regeneration.

The regeneration step relies on the pressure-dependent properties of adsorption. In particular, the adsorptive capacity per unit area of a surface increases with pressure. Typically, for small concentrations, a linear relation holds between such two quantities. At higher loads, the adsorbent approaches saturation and linearity is lost. To first order, the radon front moves along the column with a velocity proportional to the {\sl volume} flow rate. Ideally then, if the purge {\sl mass} flow rate is a fraction $f$ of the feed {\sl mass} flow rate, purging must be carried out at a pressure $P_{\rm purge}$ < $fP_{\rm feed}$ in order for the purge volume flow rate to be greater than the feed volume flow rate. Practically, radon doesn't move as a step-like front along charcoal columns but displays very pronounced tails in the concentration profile \cite{pocar-tesi:2003}; the purge pressure must then be set significantly lower than the ideal value. The performance of a filter can be described as a function of a quantity $G$ (gain) defined as 
$$
G = \Phi_{\rm purge}/ \Phi_{\rm feed},
$$
where $\Phi_{i}$ (i = purge, feed) represents the total volume flow rate through the column (proportional to the inverse of the pressure for a given mass flow rate). The bigger this ratio, the stronger the purge. It is clear how a VSA system can easily achieve very big values of $G$ with small purge fractions $f$ by purging at low pressure. 

\paragraph{Preliminary results}
Initially, a prototype PSA charcoal filter, operated between 1 and 8 bar, was run with a radon source at the inlet; it was able to significantly reduce the radon concentration at the output only with very high purge fractions (>70\%), proving extremely inefficient. It was found that nitrogen (the carrier gas and hence present in an essentially infinite amount in the system) would saturate the adsorption sites on the charcoal. This behaviour, far from linear between 1 and 8 bar, would make radon adsorption in this pressure range highly non-linear and explain the inefficiency of the device \cite{pocar-tesi:2003}. Vacuum-swing adsorption was then investigated to overcome this problem.

Studies performed on a prototype VSA device ($\sim$1.5 m$^3$/h of inlet air, 1.35 kg of charcoal per column) using a radon source at the input have, on the other hand, demonstrated radon reduction factors in excess of 10$^4$ with limited purge fractions (<10\%). The device was operated at various gains and operating pressures. Results for radon reduction are shown in fig. \ref{f:psa} (right): the data, characterized by a non-negligible spread due to uncertainties in flow rates and pressure drops, show a clear trend in the radon reduction capability as a function of gain as:
$$
{\rm log} R = -0.82 G + 0.35
$$
where $R$ is the ratio of the radon concentration at the output to that at the input of the filter. The plot shows that, for example, a gain of $\sim$4 is needed for a 1000-fold radon reduction.
Details of the device and extensive experimental results are found in \cite{pocar-tesi:2003}.

\begin{figure}
  \includegraphics[height=.28\textheight]{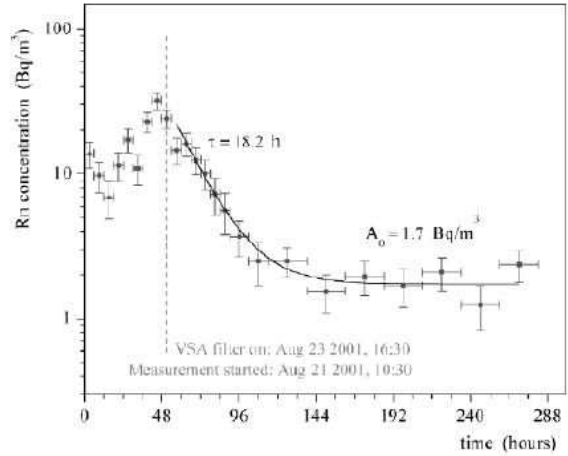}
  \caption{Radon concentration inside the Borexino clean room before and after the radon filter was turned on. The radon activity rapidly decreases at first, settling at a value $\sim$1.5 Bq/m$^3$. This value is determined by radon emanation inside the clean room, since the radon activity at the output of the filter was measured to be at least one order of magnitude smaller.}
  \label{f:rn-cr}
\end{figure}

\paragraph{Performance of the full-scale filter}

\begin{figure}
 \includegraphics[height=.33\textheight]{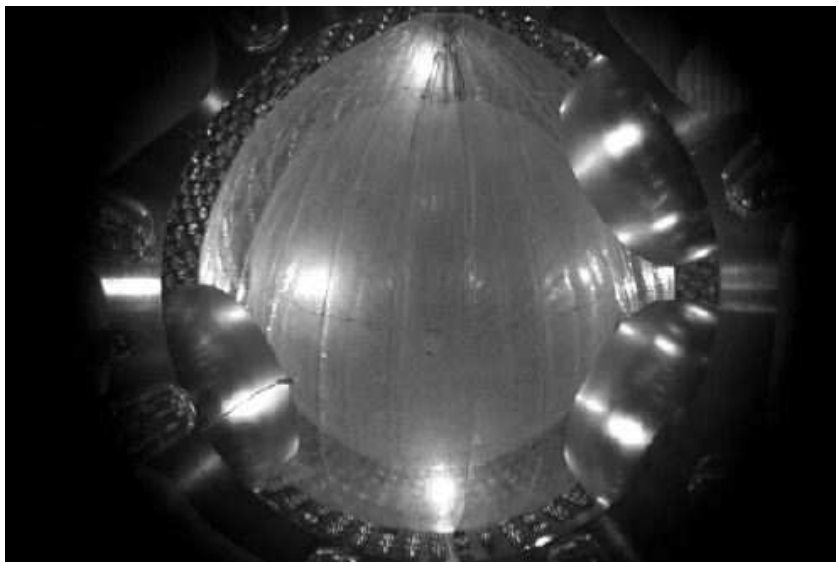}
 \caption{The Borexino nylon vessels installed and inflated inside the SSS. Ultra-low radon synthetic air carried via electroplated stainless steel pipes was used. The particulate level was better than class 1 and the radon activity was continuously monitored during inflation and was below the sensitivity limit of the detector employed of 1 mBq/m$^3$ \cite{kiko:2001}.}
 \label{f:vessels}
\end{figure}

The full-scale VSA radon filter (fig. \ref{f:scrubber}) was operated continuously during the year of vessel production (Aug. 2001 - Jul. 2002). It consisted of two charcoal columns with $\sim$250 kg of activated charcoal each. It employed a 500 m$^3$/h vacuum for the purging stage, composed of a roots blower and a rotary vane roughing pump. The inlet flow rate (dried before it reached the charcoal columns) was $\sim$100 m$^3$/h and the typical purge flow rate and pressure $\sim$12 m$^3$ and $\sim$20 mbar respectively. The gain $G$ was about 4, yielding an the expected radon reduction $\sim$1000 when scaling up the prototype results. These figures strongly depend on details of the filter such as evacuation and re-pressurization times of the columns, pressure drops across them, and residual water in the input flow, all of which would make the filter less efficient. On the other hand, the amount of charcoal is proportionally more than in the small-scale prototype, thus compensating for some of these unavoidable inefficiencies. Details of the filter design and performance are found in \cite{pocar-tesi:2003}.

The initial exponential radon reduction inside the Borexino clean room is shown in fig. \ref{f:rn-cr}. The figure also shows that the radon concentration reaches a plateau at $\sim$1.5 Bq/m$^3$. The radon concentration during the year was consistently 1-2 Bq/m$^3$; on the other hand, the radon concentration measured at the output of the VSA filter before the clean room was at least one order of magnitude smaller, at $\sim$0.1 Bq/m$^3$. This is only an upper limit since it is the sensitivity limit of the radon detector used for the measurement (Durridge RAD7 flow-through electrostatic alpha counter) and is hence compatible with the expected radon reduction stated above. 
The origin of the radon inside the clean room that limits the filter's performance is still unclear, and seems to be a recurring problem in radon-suppressed clean rooms. Small leaks ($\sim$ 5-10 m$^3$) into the clean room from the outside, maybe in places of the room with the lowest over-pressure (such as the air handling regions with high velocity flow) could explain the results. Another possible explanation is radon emanation from the HEPA filter material: this could be due to intrinsic uranium contamination of the material (that displays a very large surface area) or to dust and particulate trapped inside the filters following startup and long-term operation. In any case, it is certainly an issue worth looking into with care for future experiments; measuring the intrinsic radioactivity (specifically $^{238}$U) of HEPA filter materials could give useful insight for future applications.

\section{Vessel installation and inflation}
The Borexino nylon vessels were shipped from Princeton in Aug. 2002. Their installation, greatly delayed by a partial shut down of operations at the Gran Sasso laboratory following a scintillator spill, began in January 2004. During this time, the vessels were kept wrapped, as shown in fig. \ref{f:nesting} (right). The tight self-covering folds together with the presence of several gallons of water to keep the film moist have certainly provided a barrier against radon contamination (radon solubility in water is rather small). The extent of such protection is hard to further quantify accurately.

The Borexino SSS was operated as a class 10000 clean room until it was finally sealed (after vessel installation). The access to it was possible through a class 100 clean room, where the vessel system was partially unwrapped, hung to a horizontal rail, and carried into the SSS. It was then stretched inside the SSS by lifting it from the top polar tube assemblies. The excess film (the length of the folded vessels is $\pi$/2 times the radius of the OV) was supported on a platform mounted on a 5 m high scaffolding tower built with extremely clean material. This lower section of the vessel system was kept wrapped to prevent it from opening up and drag on the bottom of the SSS.

The two vessels were then inflated into their final spherical shape with ultra-low radon synthetic air (air instead of nitrogen was used for safety of the personnel working inside the SSS in the event of a leak from the vessels). The air was made by mixing ultra-low radon boil-off nitrogen and oxygen stored in stainless steel bottles and aged for many months underground. The synthetic air was humidified to 60\% RH with a membrane inline humidifier and fed into the vessels from the top through electroplated stainless steel pipes, part of the final Borexino fluid handling manifold. The lines were thoroughly purged with clean high pressure nitrogen until the particle count of the output flow was better than class 1. 
During vessel inflation the radon concentration of the synthetic air was continuously monitored using a large flow-through electrostatic chamber equipped with a solid state $\alpha$-counter \cite{kiko:2001}. The readings were always below the sensitivity limit of the detector of 1 mBq/m$^3$. SF$_6$ gas was added to the inlet air stream into the IV at $\sim$1\% concentration for subsequent vessel leak checking.
Once inflated, though, the most sensitive leak test for both IV and OV turned out to be given by direct measurement of the differential pressures between IV, OV, and OB, recorded with great accuracy by the pressure gauges installed in the Borexino filling lines. 
Gas leak rate limits of $\sim$0.5 cc/s for the IV and a leak rate $\sim$ 10 cc/s were achieved. These limits are better than the IV and OV design specification of 1 cc/s and 100 cc/s gas leak rates at 1 mbar over-pressure (equivalent to $\sim$0.01 and 1 cc/s liquid leak rate when the liquid to gas viscosity ratios are taken into account). A picture of the Borexino vessels installed and inflated inside the SSS, taken with one of the 7 fixed cameras of the experiment, is shown in fig. \ref{f:vessels}.

\section{Water filling the detector prior to scintillator filling}
It is foreseen to fill the Borexino detector with ultra-pure water prior to scintillator filling in order to test the Borexino filling plants. This will also provide a powerful rinsing step for all the volumes that will be in contact with the scintillator. In particular, water has been proven to be very effective at removing $^{210}$Pb from nylon surfaces. This is very important in light of possible residual radon exposure the vessels, especially the IV, could have experienced (e.g. during the long storage at Gran Sasso).

\begin{figure}
  \includegraphics[height=.24\textheight]{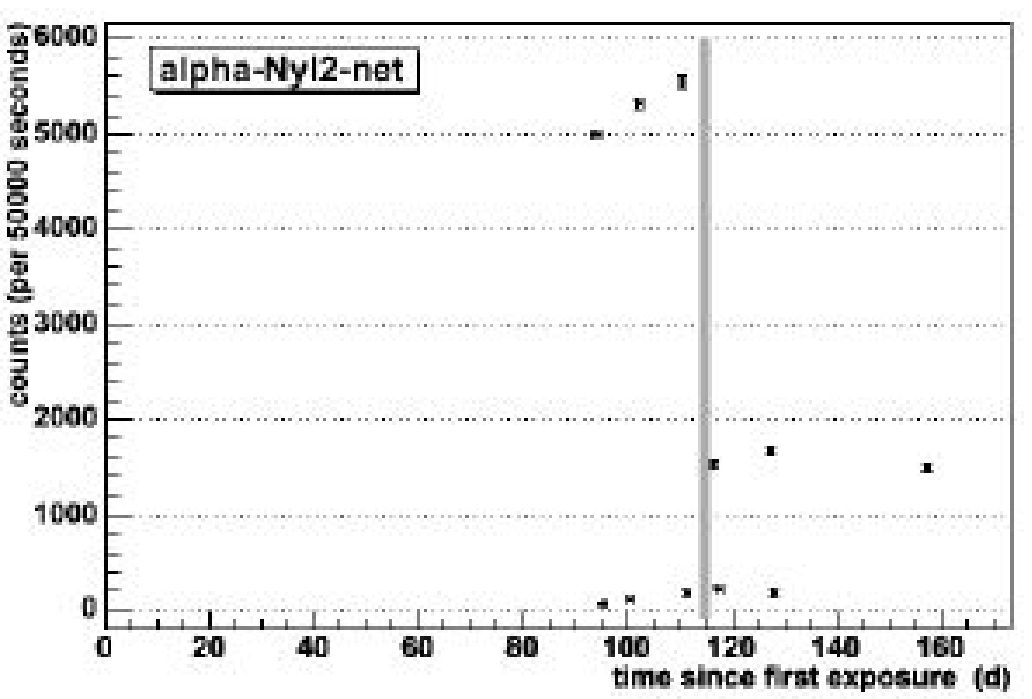}
  \hspace{.2in}
  \includegraphics[height=.24\textheight]{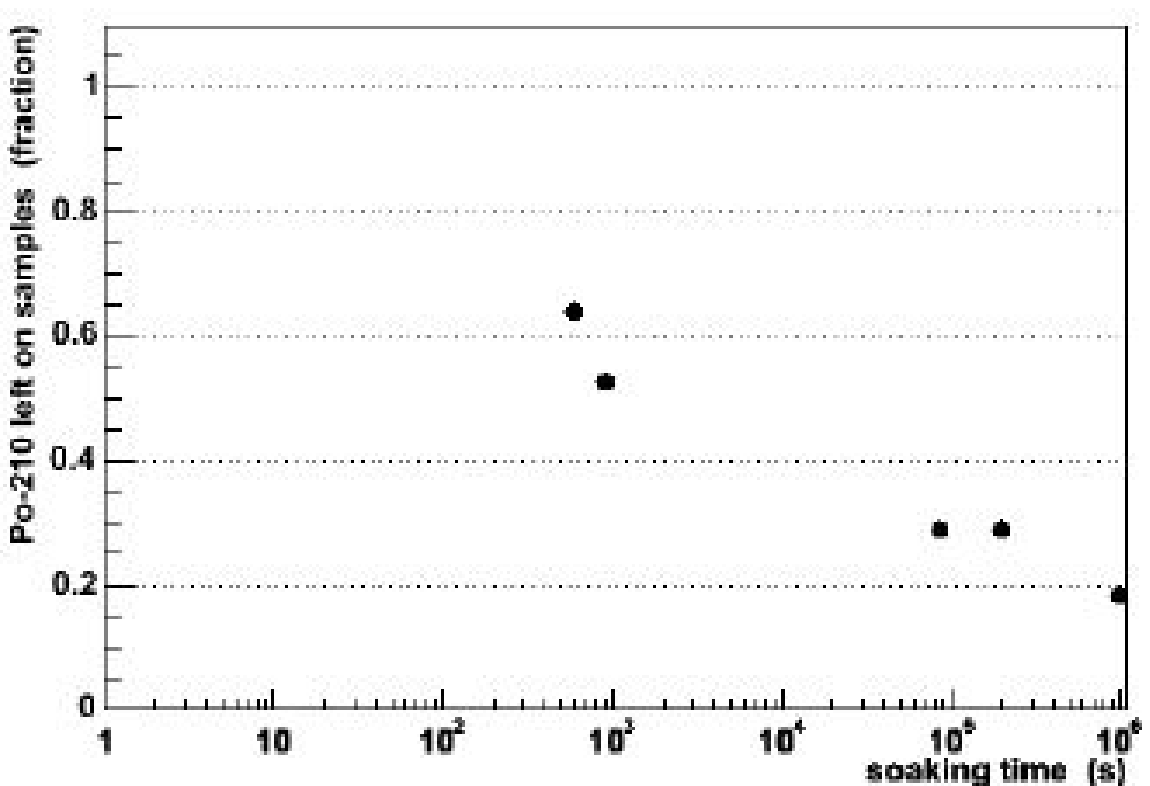}
  \caption{$^{210}$Po removal from nylon surfaces following water soaking. Left: $^{210}$Po surface activity for both surfaces of one of the samples. The vertical band is the time the sample has been immersed in DI water. The large exposure asymmetry is clearly visible in the difference in activity. It is interesting to notice that the polonium on the low activity side is essentially unchanged after immersion. Right: fraction of polonium remaining on the surface as a function of soaking time in DI water. The point around 1000 seconds corresponds to the 15-minute DI water spray-wash, similar in effect to a soak in static DI water of about twice the duration. The $^{210}$Po desorption depends on the logarithm of time; the same behaviour, with a different time constant, has also been observed in the release of $^{210}$Pb from stainless steel in water \cite{leung:2004}.}
  \label{f:polonium}
\end{figure}

Square nylon film samples (side of 4 cm) were exposed to radon-loaded air from a flow-through source for a period of $\sim$ 2 months. The samples were placed inside a stainless steel chamber (volume $\sim$10 litres) and laid out on 6 perforated steel shelves. The air was filtered by a clean room grade filter to mimic the particulate contamination during vessel fabrication. The nylon samples were cut out from the left over material used for the IV (Sniamid, see previous sections). 
The 46 keV $\gamma$-ray activity from $^{210}$Pb, as well as the 5.3 MeV $\alpha$ activity from $^{210}$Po were measured with an intrinsic germanium and a silicon surface barrier detector respectively. The activities of both isotopes were measured just before and just after the samples were immersed for a given time (10 minutes to 10 days) in DI water (1 litre, $\rho$ > 16 M$\Omega \cdot$cm). One sample was spray-washed for $\sim$15 minutes with running DI water.
The $^{210}$Po activity, that had not reached equilibrium yet with $^{210}$Pb due to its long 200-day lifetime, was measured a few times at different times prior to water soaking to account for the buildup rate. The main results from these measurements were:
\begin{enumerate}
\item[a.] essentially all $^{210}$Pb is removed after 1 day of soaking. A 10-minute soak washes off $\sim$75\% of the lead. There is some indication that 5-10\% of the lead is not removed even after 10 days, but this result has a large error due to the very low gamma activity of the samples;
\item[b.] $^{210}$Po is released much more slowly than lead. Even after 10 days, approximately 20\% of the activity is still on the surface (see fig. \ref{f:polonium}, right).
\item[c.] there was a significant asymmetry of activity between the two sides of the samples. The side directly facing the input flow inside the exposure chamber had a much higher (10-100 times) polonium activity than the other, facing the perforated metal shelves. It was not possible to directly measure this asymmetry in the lead activity since the gamma rays would go through the 125 $\mu$m sample. 
\end{enumerate}
These results strongly support water filling the Borexino detector. Soaking the vessels in ultra-pure DI water for a period of the order of one month would dramatically reduce the residual $^{210}$Pb and $^{210}$Po contamination on their surface. The real need for such procedure depends on the actual radon exposure of the vessels during fabrication and installation and on the dynamics of lead and polonium release from nylon surfaces into the organic scintillator. In any case, an aggressive spray-wash of the IV, which was considered as an option, is proven unnecessary.
Other interesting information can be extracted from the data. The striking asymmetry in $^{210}$Po activity is presumably related to the lead deposition mechanism. There could be a correlation between the observed residual lead activity and the polonium activity on the low activity side of the samples, unchanged by soaking in DI water, that could be due to different levels of isotope implantation in the material. Further studies of this sort could help better understand radon daughter deposition on surfaces in clean room environments that could be very valuable for many low background experiments.

%%%%%%%%%%%%%%%%%%%%%%%%%%%%%%%%%%%%%%%%%%%%%%%%
%% BACKMATTER
%%%%%%%%%%%%%%%%%%%%%%%%%%%%%%%%%%%%%%%%%%%%%%%%

\begin{theacknowledgments}
The results reported in this paper were possible thanks to many years of R\&D carried out within the Princeton/Borexino group. The installation and inflation phase at Gran Sasso saw the participation of many other members of the Borexino collaboration who contributed to make the commissioning of the nylon vessels successful. Some crucial low radioactivity measurement of the film were performed by the Heidelberg/Borexino group, also responsible for the supply of ultra-low radon synthetic air and nitrogen at LNGS. 
\end{theacknowledgments}



\bibliographystyle{aipproc}   % if natbib is available

\bibliography{mybib}

\IfFileExists{\jobname.bbl}{}
 {\typeout{}
  \typeout{******************************************}
  \typeout{** Please run "bibtex \jobname" to optain}
  \typeout{** the bibliography and then re-run LaTeX}
  \typeout{** twice to fix the references!}
  \typeout{******************************************}
  \typeout{}
 }

\end{document}